\documentstyle[12pt]{article}
\def\be{\begin{equation}}
\def\ee{\end{equation}}
\begin{document}
\setcounter{footnote}0
\begin{center}
\hfill AIV - 98/I\\
\vspace{0.3in}
\bigskip
\bigskip
\bigskip
{\Large \bf  Domain Walls in MQCD\\
\bigskip
and\\
\bigskip
Monge-Amp\`{e}re
Equation
 }
\date{}
\bigskip
\bigskip
\bigskip
\bigskip
\bigskip
\bigskip
\\{\Large Anastasia Volovich}
\bigskip
\begin{center}
{\it LPTHE,  Universit$\acute{e}$s Paris VI - Paris VII\\4 place Jussieu,
F-75252 Paris Cedex 05, France\\
nastya@lpthe.jussieu.fr, nastya@itp.ac.ru \footnote{On leave from  Moscow
State University
and L. D. Landau Institute for Theoretical Physics, Moscow, Russia.}}
\end{center}
\bigskip

\end{center}
\begin{abstract}

We study Witten's proposal that a domain
wall exists in M-theory fivebrane version of QCD (MQCD) and
that it can be represented as a supersymmetric three-cycle
in $G_{2}$ holonomy manifold. It is shown that equations
defining the U(1) invariant domain wall for $SU(2)$  group
can be reduced to
the Monge-Amp\`{e}re equation. A proof of an algebraic formula
of Kaplunovsky, Sonnenschein
and Yankielowicz is presented. The formal solution  of
equations for domain wall is constructed.

\end{abstract}
\newpage

\section{Introduction}

Starting from the pioneering work by Hanany and Witten \cite{HW}
the study of the low-energy dynamics of a certain class of intersecting
brane
configurations has shed much light on non-perturbative properties of
gauge theories \cite{HW}-\cite{itep}. Recently,
Witten explored the minimal  $N=1$
model with an $SU(n)$ vector multiplet in four dimensions \cite{Witten}.
He showed how for this model some of the outstanding properties of the
ordinary QCD such as confinement, a mass gap and spontaneous breaking of
a discrete chiral symmetry can be approached from M-theory point of view.
The consideration of N=1 gauge theory in the geometric engineering
approach was performed in \cite{BersVafa}.

To describe the M5-brane version of QCD (MQCD) \cite{Witten},
one  starts from the brane configuration in type IIA
superstring theory with space-time coordinates $(x^{0},$
$x^{1},$$...,x^{9})$
and studies a configuration arising from $n$ D4-branes suspended between
two NS5-branes located at $x^{6}=0$ and $x^{6}=S_{0}$
\cite{EGK}, \cite{Witten}.
D4-branes world-volumes occupy
$(x^{0},x^{1},x^{2},x^{3},x^{6})$, with $0 \leq x_{6} \leq S_{0}$,
NS5-brane's world-volume is spanned by
$(x^{0},x^{1},x^{2},$ $x^{3},x^{4},x^{5})$ and another NS'5-brane's by
$(x^{0},x^{1},x^{2},x^{3},x^{7},x^{8})$,
where $S_{0}$ is an arbitrary length.
Then the world-volume theory
on D4s is four dimensional Super Yang Mills with gauge group
$SU(n)$ and $N=1$ supersymmetry.
Elevating to the M-theory picture by adding the coordinate $x^{10}$
makes possible a solution of the theory
as follows \cite{Witten}. Reinterpreted as a brane configuration
embedded in eleven dimensional spacetime,
the entire brane configuration corresponds to a
single smooth M5-brane with world-volume $R^{1,3} \times \Sigma$,
where $\Sigma$ is a Riemann surface, embedded in three-dimensional
space $Y$ with coordinates $\upsilon, w, t=e^{-s}$
as $ {\upsilon} w=\zeta,~{\upsilon}^{n}=t$, here
$\upsilon=x^{4}+ix^{5},~w=x^{7}+ix^{8},~s=x^{6}+ix^{10}, 0 \le x^{10}
\leq 2\pi$
and $\zeta$ is a complex constant.
Analyzing the symmetries, one can notice \cite{Witten}
that $Z_{n}$ symmetry:
$t \to t,~\upsilon \to \upsilon, w \to e^{2\pi i/n} w$
is only symmetry at infinity, which doesn't leave the first
equation defining $\Sigma$ invariant. Thus \cite{Witten},
this symmetry is spontaneously broken and the theory has $n$
distinct vacua, specified by the curves
$w=exp({2 \pi i}/{n}) \zeta {\upsilon}^{-1}, t={\upsilon}^{n}$.

 A consequence of the spontaneously broken chiral symmetry
is that there can be a domain wall separating different vacua.
BPS--saturated domain walls in four dimensional
supersymmetric gauge theories have been considered
in \cite{Shifman},\cite{KSS}.
Witten has suggested that a BPS--
saturated domain wall exists in MQCD and that it can be represented
as a supersymmetric three-cycle in the sense of Becker {\it et al}
\cite{Becker1},\cite{Becker} with a prescribed
asymptotic behavior.
The domain wall is described \cite{Witten} as an M-theory
fivebrane of the form $ {\bf {R}}^{3} \times \cal S$, where $ {\bf
{R}}^{3} $
is parameterized by $x^{0}, x^{1}, x^{2} $ and $\cal S$ is
a three-surface in the seven manifold $ \tilde{Y} = {\bf {R}} \times Y$,
here ${\bf R}$ is the copy of $x^{3}$ direction.
Near $x^{3}=- \infty $, $\cal S $ should look like
${\bf R} \times \Sigma$,
where $\Sigma $ is the Riemann surface defined by
$w = \zeta {\upsilon }^{-1}$, $t={\upsilon}^{n}$.
Near  $x^{3}=+\infty,~\cal S $ should look like
${\bf R} \times {\Sigma}^{'}$, where ${\Sigma}^{'}$
is the Riemann surface of an "adjacent" vacuum,
defined by
$w=exp({2 \pi i}/{n}) \zeta {\upsilon}^{-1}, t={\upsilon}^{n}. $
MQCD is by no means identical to QCD, it depends on one extra
parameter -- type IIA string coupling constant.
For the domain wall to be in the
universality class of SQCD, $\cal S$ must be invariant under U(1)
symmetry $t \to e^{in\delta} t,$ $\upsilon \to e^{i \delta} \upsilon,$
$w \to e^{-i\delta} w.$
Different approaches to the problem of
domain walls in MQCD have been explored in \cite{AV,Fayya,Scern,KSY}.
Equations defining the domain wall have been derived and studied
in \cite{AV}.

The aim of this note is the consideration of such $U(1)$
symmetric  $\cal S$
which is a supersymmetric three-cycle
in $\tilde{Y}$ with the described asymptotic behavior.
We use  Witten's U(1) invariant ansatz for $SU(2)$ group and
an algebraic formula of Kaplunovsky, Sonnenschein
and Yankielowicz (KSY). We consider two
gauges when one of equations has the form of conservation low.  This
permits
to reduce the system of equations to one equation.  We show that the
equations defining the domain wall can be reduced to the Monge-Amp\`{e}re
equation. A proof of the KSY formula is presented.
This formula is very useful
for the consideration of domain walls in MQCD. The formal
solution of equations for the domain wall is constructed
using a special separation of variables for group $SU(2)$
in the spirit of \cite{AV}.

\section{Supersymmetric Cycles in Various Dimensions}

 A supersymmetric cycle is defined by the property that the
world-volume theory of a brane wrapping around it is supersymmetric.
To study supersymmetric cycles one uses the concept
of calibration \cite{Becker}, t.e.  a closed $p$-form $\tilde{\Phi}$ on
a Riemannian manifold of dimension $n$ such that $\tilde{\Phi}$
has comass 1.
Submanifolds for which there is equality are said to be calibrated by
$\tilde{\Phi}$. The calibrated submanifold has
the least volume in its homology class. This
provides a natural geometrical interpretation of the BPS bound for
D-branes wrapped
around such submanifolds, with the calibrated submanifolds
corresponding to BPS-states, which saturate the bound.

The conditions for the supersymmetric cycles in Calabi-Yau
3-folds have been analyzed in \cite{Becker1}. It was shown that
a supersymmetric three-cycle is one for which the pullback of
K\"{a}hler form  $J$ vanishes and the pullback of the holomorphic
3-form $\Omega$ is a constant multiple of volume element, namely
$\ast X(J) = 0,~~\ast X(\Omega) \sim 1$,
where $X(.)$ denotes the pullback and $\ast$ is a Hodge dual
on membrane world-volume.

In the case of domain walls in MQCD \cite{Witten} one deals with
a seven dimensional flat manifold $\tilde{Y}$ of $G_{2}$ holonomy and with
the
associative calibration $\tilde{\Phi}$. The group $G_2$ is most naturally
defined
as
the automorphism group of the octonions or Cayley numbers
${\bf O}={\bf H}(+)$, the eucledian algebras obtained from the
quaternions by Cayley-Dickson process \cite{SandC}.
If we choose
the local veilbein
so that the metric on $\tilde{Y}$
is $\sum_{i=1}^{n} e_{i} \otimes e_{i},$
locally the $G_{2}$ invariant 3-form $\tilde{\Phi}$ can be written as
\cite{Vafa}
$$
\tilde{\Phi}=e_{1} \wedge e_{2} \wedge e_{7}
+ e_{1} \wedge e_{3} \wedge e_{6}
+ e_{1} \wedge e_{4} \wedge e_{5}
+ e_{2} \wedge e_{3} \wedge e_{5}-$$
\be
e_{2} \wedge e_{4} \wedge e_{6}
+ e_{3} \wedge e_{4} \wedge e_{7}
+ e_{5} \wedge e_{6} \wedge e_{7}.
\label{Phi}
\ee
A supersymmetric three-cycle $\cal S$ in $\tilde{Y}$ is one for which
the
pullback  of this three-form is a constant multiple of
the volume element \cite{Becker}.
The invariant forms are related by the dimensional
reduction. If we set

\be
e_1=dx^{10},~~e_2=dx^{5},~~e_3=dx^{3},
~e_4=dx^{7},~~e_5=dx^{4},~~e_6=dx^{6},~~e_7=dx^{8},
\label {emb}
\ee
then the form $\tilde{\Phi}$ can be written as
\be
\tilde{\Phi} = Im(\Omega) + \frac{i}{2}dx^{3} \wedge J,
\label{supercyc}\ee
where
\be
\Omega=dv\wedge dw\wedge dt/t
\label{omega}
\ee
and
\be
J=dv\wedge d{\bar v}+dw\wedge d{\bar w} +dt\wedge d{\bar t}/|t|^2
\label{CY}
\ee
are K\"{a}hler and holomorphic forms in Calabi-Yau 3-fold Y.
If one equates the pullback of the $J$ to zero, then
from the condition for supersymmetric cycle in 7 dimensional manifold
one gets the condition in 6 dimensional manifold.
This is probably a relation between the equations considered here
and in the recent paper \cite{Fayya}, where another approach to the
problem of
the domain wall in MQCD was suggested.

Baulieu, Kanno, and Singer have developed an almost topological
theory, so called BRSTQFT in 8 dimensions \cite {BKS1, BKS2}. It seems
that supersymmetric cycles in various dimensions can be obtained
by the dimensional reduction from the BRSTQFT.

\section{ Equivalent Form of Supersymmetric Cycles}

We will be looking for a supersymmetric three-cycle $\cal{S}$
with worldvolume coordinates $(y_1,y_2,y_3)$
in a 7-manifold $\tilde{Y}$ with coordinates
$(x^{3}, \upsilon,w,t)$ which near  $x^{3}=-\infty$
looks like
${\bf R} \times \Sigma$,
where $\Sigma $ is the Riemann surface defined by
$w = \zeta {\upsilon }^{-1}$, $t={\upsilon}^{n}$ and
near  $x^{3}=+\infty,~\cal S $ like
${\bf R} \times {\Sigma}^{'}$, where ${\Sigma}^{'}$
is the Riemann surface of an "adjacent" vacuum,
defined by
$w=\zeta e^{2 \pi i/n} {\upsilon}^{-1}, t={\upsilon}^{n}. $
In this note  we will consider group $SU(2)$.

Let us make an embedding of $\cal{S}$ into $\tilde{Y}$

\be
\upsilon=z_1(y_1,y_2,y_3),
\label{gv}
\ee

\be
w=z_2(y_1,y_2,y_3),
\label{gw}
\ee

\be
s=z_3(y_1,y_2,y_3)
\label{gt}
\ee

\be
x^{3}=y_3
\ee
and introduce the complex 3-vectors $a_k$ with components

\be
a^i_{k}=
\frac{\partial z_i}{\partial y_k},
\label{ai}
\ee
where $i, k =1,2,3.$

The condition for $\cal{S}$ to be a supersymmetric 3-cycle
in these notations is \cite{Becker1,Becker,AV}

\be
\sqrt{det ||h_{mn}||} dy_1 \wedge dy_2 \wedge dy_3 = \tilde{\Phi}
~~or~~~
det ||h_{mn}||=\Phi^2,
\label{sucyc}
\ee
where
\be
h_{mn}=Re(a_m^{*} \cdot a_n )+ \delta_{m3}\delta_{n3}
\label{h}
\ee
is an induced metric
and

\be
\Phi=Im [(a_1^{*} \cdot a_2)- (a_{3} \cdot a_{1}\times a_{2})]
\label{psie}
\ee
is a pullback of $G_2$ invariant form and $(a\cdot b)= \sum _{i=1}
^{3}a^i b^i$.

Due to Theorem  proved in Appendix
we have
\be
det ||h_{mn}||=\Phi^2 +|R|^2 +Re^2(a_3\cdot R),
\label{lemma}
\ee
where
\be
R=a_1\times a_2 +\frac{i}{2}\epsilon_{mnk}a_m^{*} Im(a_n^{*} \cdot a_k)
\label{R}
\ee
and the requirement for the surface $\cal{S}$ to be a supersymmetric three
cycle (\ref{sucyc}), is reduced to the
equation derived by
Kaplunovsky, Sonnenschein and Yankielowicz \cite{Scern,KSY}

\be
R=0,
\label{R0}
\ee
or
\be
a_1\times a_2 +\frac{i}{2}\epsilon_{mnk} a_m^{*} Im(a_n^{*} \cdot a_k)=0.
\label{EQ}
\ee

Let us note the following

{\bf Proposition. } The relation
\be
ia^*_3 Im (a^*_1 \cdot a_2 )
+a_1 \times a_2 =0
\label{pro1}
\ee
implies (\ref{EQ}).

\section{U(1) Ansatz for  Domain Wall}

Let us consider the group $SU(2)$ and make an embedding:

\be
\upsilon=z_1=f(y_1,y_3)e^{iy_2},
\label{v}
\ee

\be
w=z_2=g(y_1,y_3)e^{-iy_2},
\label{w}
\ee

\be
s=z_3=-h(y_1,y_3) - 2iy_2.
\label{t}
\ee
Under this  U(1) invariant ansatz the equation
(\ref{EQ})
\be
-ia_1\times a_2 +a_1^* Im(a_2^* \cdot a_3)+a_2^* Im(a_3^* \cdot a_1)+
a_3^* Im(a_1^* \cdot a_2)=0,
\label{son}
\ee
where

$$a_1=(\partial_1 f \cdot e^{iy_2},
\partial_1 g \cdot e^{-iy_2},-\partial_1 h),$$

$$
a_2=(i f \cdot e^{iy_2},-i g\cdot e^{-iy_2},-2i),$$

\be
a_3=(\partial_3 f\cdot e^{iy_2},
\partial_3 g\cdot e^{-iy_2}, -\partial_3 h)
\label{aie}
\ee
is reduced to  the following equations for
complex functions $f,~g,~h:$

\begin{equation}
\{K, f^{*}\}_{(3,1)}-
iPf^{*}-2(2  \partial_{1} g+ g\partial_{1}h)=0;
\label{e1}
\end{equation}

\begin{equation}
\{K, g^{*}\}_{(3,1)}+
iPg^{*}+2(2  \partial_{1} f-f\partial_{1}h)=0;
\label{e2}
\end{equation}

\begin{equation}
-\{K, h^{*}\}_{(3,1)}+2
iP-2 \partial_{1} (fg)=0,
\label{e3}
\end{equation}
where

$$P=-Im[\{f,f^{*}\}_{3,1}+\{g,g^{*}\}_{3,1}+\{h,h^{*}\}_{3,1}],~~~~
K=|g|^2-|f|^2-2h-2h^{*}$$
and the Poisson brackets are defined as

$$\{g,f\}_{(i,j)}=\partial_i g\partial_j f-
\partial_j g \partial_i f,~~~i,j=1,.3.$$

The  boundary conditions for group $SU(2)$ read:
\be
fg|_{y_{3}=\mp \infty}=\pm\zeta,~~~~~
f^2 e^{-h}|_{y_{3}=\pm \infty}=1, ~~~~~Im \zeta =0.
\label{bc}
\ee

Note that we have 3 complex equations for 3 complex functions,
but not all of them are independent. From equation (\ref{e3})
and its complex conjugated one gets the following  equation
\begin{equation}
\{K,h+ h^{*}\}_{(3,1)}+4\partial_{1} Re(fg)=0,
\label{e3c}
\end{equation}
which in fact follows from (\ref{e1}) and (\ref{e2}).

\subsection{Real Functions}

Let us now assume   that functions $f, ~g,~h$ are real.
Then $P=0$ and equations (\ref{e1})-(\ref{e3})
are reduced to the following equations

\begin{equation}
\{K, f\}_{(3,1)}-
2(2  \partial_{1} g+ g\partial_{1}h)=0,
\label{e1r}
\end{equation}
\begin{equation}
\{K, g\}_{(3,1)}+2(2\partial_{1} f-f\partial_{1}h)=0,
\label{e2r}
\end{equation}
\begin{equation}
\{K, h\}_{(3,1)}+2 \partial_{1} (fg)=0,
\label{e3r}
\end{equation}
with
\be
K=g^2-f^2-4h.
\label{K}
\ee
Equation (\ref{e2r}) is the
combination of equations (\ref{e1r}) and (\ref{e3r}), thus it can be
dropped
out
and we are left with the following system of equations:
\be
\{4h-g^2,f\}_{(1,3)}-4\partial _1 g -2g\partial _1 h=0,
\label{1eq}
\ee
\be
\{g^2-f^2,h\}_{(1,3)}-2\partial _1 (fg)=0.
\label{3eq}
\ee

\subsection {Formal Solution}

To check a self consistency of the above equations let us derive
a formal
solution of these equations for group SU(2).
The ansatz for functions $f, ~g,~h$ considered here is of the form:

\begin{equation}
f(y_1, y_3)=f_{0}(y_1)+\sum_{k=1}^{\infty} {\gamma}^{2k}(y_3) f_{2k}(y_1),
\label{af}
\end{equation}

\begin{equation}
g(y_1, y_3)=-\zeta
\beta(y_3)  \cdot (g_{0}(y_1)+\sum_{k=1}^{\infty} {\gamma}^{2k}(y_3)
g_{2k}(y_1)),
\label{ag}
\end{equation}

\begin{equation}
h(y_1, y_3)=h_{0}(y_1)+\sum_{k=1}^{\infty} {\gamma}^{2k}(y_3) h_{2k}(y_1),
\label{ah}
\end{equation}
where
$$\gamma(y_3)=\frac{1}{e^{y_3}+e^{-y_3}}=\frac{1}{2\cosh y_3},$$

$$\beta(y_3)=\frac{e^{y_3}-e^{-y_3}}{e^{y_3}+e^{-y_3}}=
\tanh y_3.$$

We notice that for the above ansatz the boundary
conditions
are trivially satisfied if
$ f_0 g_0=1,$ and $f_0 ^2 e^{-h_0}=1$.

Due to the simple differential algebra
$\partial_3 \beta =4 \gamma^2, ~
\partial_3 \gamma^{k}=-k \beta \gamma^k, ~ \beta^2=1-4 \gamma^2$
the equations (\ref{1eq}-\ref{3eq}) will take the form
of the equations on $f_{2k},~g_{2k},~h_{2k}$:
\be
-\zeta g_0 \partial_1 f_{2k}+
(2k f_0 \partial_1 h_0-\zeta \partial_1 g_0)f_{2k}
-\zeta f_0 \partial_1 g_{2k}
-(2k \zeta^2 g_0 \partial_1 h_0+\zeta \partial_1 f_0)g_{2k}+
\ee
$$
k(2 \zeta^2 g_0 \partial_1 g_0-2 f_0 \partial_1 f_0)h_{2k}=
L_{2k},
$$
\be
k(2 \zeta^2 g_0 \partial_1 g_0 -4 \partial_1 h_0)f_{2k}+
2\zeta \partial_1 g_{2k}-
(2k \zeta^2 g_0 \partial_1 f_0 - \zeta \partial_1 h_0)g_{2k}+
\ee
$$
\zeta g_0 \partial_1 h_{2k}+
4k \partial_1 f_0 h_{2k}=M_{2k},
$$
where
\be
L_{2k}=
\zeta \partial_1(f_{2k-2m}g_{2m})-
m h_{2m}K_{2k-2m}^{(1)}-
\partial_1 h_{2k-2m} K_{2m}^{(3)}-
\partial_1 h_0 K_{2k}^{'(3)},
\ee
\be
M_{2k}=-\zeta g_{2k-2m}\partial_1 h_{2m}-
\partial_1 f_{2k-2m} K_{2m}^{(3)}-
mf_{2m} K_{2k-2m}^{(1)}-
\partial_1 f_0 K_{2k}^{'(3)}
\ee
\be
K_{2k}^{(1)}=\tilde{K}_{2k}^{(1)}+K_{2k}^{'(1)},
~~
K_{2k}^{(3)}=\tilde{K}_{2k}^{(3)}+K_{2k}^{'(3)},
\label{KK}
\ee
\be
\tilde{K}_{2k}^{(1)}=
2 \zeta^2 \partial_1 (g_0 g_{2k})
-2 \partial_1 (f_0 f_{2k})-
4 \partial_1 h_{2k},
~
\tilde{K}_{2k}^{(3)}=
-2k \zeta^2 g_0 g_{2k}+
2k  f_0  f_{2k}+
4k  h_{2k},
\ee
\be
K_{2k}^{'(1)}=
-8 \zeta^2 (\delta_{k1}g_0 \partial_1 g_0+
\partial_1 (g_0 g_{2k-2}))+
\ee
$$
\sum_{m=1}^{k-1}
2 \zeta^2 g_{2k-2m}\partial_1 g_{2m}
-8 \zeta^2 g_{2k-2m-2}\partial_1 g_{2m}
-2 f_{2k-2m}\partial_1 f_{2m} ,
$$

\be
K_{2k}^{'(3)}=
g_{0} \zeta^2
(4 g_0 \delta_{k1}+
4(2k-1)g_{2k-2})-
\ee
$$
\sum_{m=1}^{k-1}
\zeta^2 g_{2k-2m}
(2m g_{2m}-4g_0 \delta_{m1}-
4(2m-1)g_{2m-2})+
2m f_{2k-2m} f_{2m}.
$$

We notice that we have two equations for three
functions, so it seems useful to
assume that for example $h_{2k}=0.$
Let us also
take a parametrization of the form
\be
h_0=y_1,~~f_0=e^{y_1/2},~~g_0=e^{-y_1/2}.
\ee
Then the equations will take the form:
\be
-\zeta e^{-y_1/2}\partial_1 f_{2k}+
(2k e^{y_1/2}+\frac{\zeta}{2}e^{-y_1/2})f_{2k}
-\zeta e^{y_1/2}\partial_1 g_{2k}-
(2k \zeta^2 e^{-y_1/2}+\frac{\zeta}{2}e^{y_1/2})g_{2k}=L_{2k}
\label{Pe1}
\ee
\be
k(-\zeta^2 e^{-y_1}-4)f_{2k}
+2\zeta \partial_1 g_{2k}-
(k \zeta^2 -\zeta)g_{2k}=M_{2k}
\label{Pe2}
\ee
For example, in the first order
one can easily get
\be
-\zeta e^{-y_1/2} \partial_1 f_{2}+
(2 e^{y_1/2}+\frac{\zeta}{2} e^{-y_1/2})f_2
-\zeta e^{y_1/2}\partial_1 g_2-
(2\zeta^2 e^{-y_1/2}+\frac{\zeta}{2}e^{y_1/2})g_2=
-4 \zeta^2 e^{-y_1/2},
\label{2o1e}
\ee
\be
-(\zeta^2 e^{-y_1}+4)f_2+
2 \zeta \partial_1 g_2-
(\zeta^2-\zeta)g_2=-2 \zeta^2 e^{-y_1/2}.
\label{2o2e}
\ee
From equation (\ref{2o2e}) one can express $f_2$,
substitute this expression to (\ref{2o1e})
and get one equation on $g_2$, which can be solved in quadratures.
Similarly, for the $k^{th}$ order.
So we get the following

{\bf Proposition.}
There exists a solution of equations (\ref{1eq}) and (\ref{3eq})
of the following form
\begin{equation}
f(y_1, y_3)=e^{y_1/2}+\sum_{k=1}^{\infty} {\gamma}^{2k}(y_3) f_{2k}(y_1),
\label{Pf}
\end{equation}

\begin{equation}
g(y_1, y_3)=-\zeta
\beta(y_3)  \cdot (e^{-y_1/2}+\sum_{k=1}^{\infty} {\gamma}^{2k}(y_3)
g_{2k}(y_1)),
\label{Pg}
\end{equation}

\begin{equation}
h(y_1, y_3)=y_1,
\label{Ph}
\end{equation}
where $f_{2k}$ and $g_{2k}$ satisfy the recursive relations
(\ref{Pe1}) and (\ref{Pe2}). The solution (\ref{Pf})-(\ref{Ph})
satisfies
the boundary conditions (\ref{bc}).

\smallskip

{\it Remark. Small $\zeta$}

Let us take $\zeta \to 0$.
As it was pointed in \cite{Witten} $\zeta \to 0$
and $R \to \infty$ corresponds
to the small QCD scale and one gets the ordinary super Yang-Mills.

In this case the equations
will take the form
\be
2ke^{y_1/2}f_{2k}=-2m f_{2k-2m}f_{2m},
\ee
\be
-4kf_{2k}=0,
\ee
so the solution is $f_{2k}=0$ and $h_{2k}$
is arbitrary.
So, in principle, one can take $h_{2k}=0$
then the surface
\be
\upsilon w=-\zeta tanh(x_3),~~~t=v^2
\ee
will correspond to the domain wall
for small $\zeta$.

From (\ref{e3r}) one notices that if $K=y_1$ or $h=y_1$  then this
equation
becomes the conservation law. We consider these two cases separately.

\subsection{Monge-Amp\`{e}re equation}

Let us first take a parameterization  of the surface
defined by (\ref{e1r})-(\ref{e3r}) as

\be
h=y_1.
\ee
This ``gauge" was considered also in \cite{Scern,KSY}. Then we get

\begin{equation}
4 \partial _3 f+\{g^2, f\}_{(3,1)}-
2(2  \partial_{1} g+ g)=0
\label{e1g}
\end{equation}

\begin{equation}
4\partial _3 g-\{f^2, g\}_{(3,1)}+
2(2  \partial_{1} f-f)=0
\label{e2g}
\end{equation}

\begin{equation}
\partial _3(g^2-f^2) + \partial_{1} (2fg)=0.
\label{e3rg}
\end{equation}
Equation (\ref{e3rg}) has the form
of a conservation law.
From this equation it follows that there exists a function $\chi$
such that

\be
fg=-\frac{1}{2}\partial _3 \chi,
\label{chi11}
\ee

\be
g^2-f^2=\partial _1 \chi.
\label{chi21}
\ee
One can express $g^2$  and $f^2$  in term of $\chi$ as

\be
g^2=\frac{1}{2}(\partial _1 \chi +\sqrt{(\partial _1 \chi )^2
+(\partial _3 \chi)^2  }),
\label{gg21}
\ee

\be
f^2=\frac{1}{2}(-\partial _1 \chi +\sqrt{(\partial _1 \chi )^2
+(\partial _3 \chi)^2  }),
\label{gf21}
\ee

\be
f^2+g^2=\sqrt{(\partial _1 \chi )^2
+(\partial _3 \chi)^2 }.
\label{gg+f1}
\ee

Multiplying (\ref{e1g}) on $g$ and (\ref{e2g})  on $f$  and sum up
we get

\begin{equation}
4\partial _3 (fg)+\{g^2-f^2, fg\}_{(3,1)}-
2\partial_{1} (g^2-f^2)- 2(f^2+g^2)=0.
\label{sum}
\end{equation}

Substituting (\ref{chi11}), (\ref{chi21}) and (\ref{gg+f1})
in (\ref{sum}) one gets
\begin{equation}
- \partial ^2 _1 \chi- \partial^2 _3\chi +
\frac{1}{4}\partial ^2 _1 \chi\cdot \partial^2 _3\chi -
\frac{1}{4}(\partial ^2 _{13} \chi)^2
- \sqrt{(\partial _3 \chi )^2
+(\partial_1 \chi)^2}=0.
\label{MMA}
\end{equation}

One can easily derive the boundary conditions for function
$\chi$ from (\ref{bc})

\be
\partial _3\chi|_{y_3 \to \pm\infty}=\pm 2\zeta,~~~~~
\partial _1\chi|_{y_3 \to \pm\infty}=\zeta^2 e^{-y_1}-e^{y_1}=
-2e^{\xi}\sinh (y_1-\xi).
\ee
where $\zeta=e^{\xi}$.
Equation (\ref{MMA}) is in fact a Monge-Amp\`{e}re equation
\cite{Pogor,TA,Yau}.
One can write it in the canonical form if one sets
$\phi(x,y)=\frac{1}{4} \chi(y_1,y_3).$ Then (\ref{MMA})
reads
\begin{equation}
\phi_{xx}\phi_{yy}-\phi_{xy}^2=
\phi_{xx}+ \phi_{yy}+
\sqrt{\phi_x^2 +\phi_y^2}.
\label{MoAm}
\end{equation}

We have to find a solution of (\ref{MoAm}) in the plane with the following
boundary conditions:

\be
\lim_{y \to \pm\infty} \phi_x=\frac{1}{4}(\zeta^2 e^{-x}-e^{x}),
~~~
\lim_{y \to \pm\infty} \phi_y=\pm \frac{1}{2}\zeta.
\label{bcMA}
\ee
Here $\zeta $ is a real parameter.

Notice vacuum solutions of (\ref{MoAm})

\be
\phi =-\frac{e^{\xi}}{2}(\mp y+ \cosh(x-\xi)).
\ee

Let us recall that the general
Monge-Amp\`{e}re equation has the form \cite{Pogor}

\begin{equation}
\phi_{xx}\phi_{yy}-\phi_{xy}^2=
a\phi_{xx}+2b\phi_{xy}+c\phi_{yy}+ g.
\label{Pog}
\end{equation}
where $a,b,c $ and $g$ are functions of
$x,y,\phi,\phi_{x}$ and $\phi_{y}$.
In our case $a=c=1$, $b=0$ and
$g=\sqrt{\phi_x^2 +\phi_y^2}.  $

The Monge-Amp\`{e}re equation (\ref{Pog}) is called
strongly elliptic if $g>0$ and the quadratic form
$a \rho ^2+2b\rho \eta +c\eta ^2$ is non-negatively defined \cite{Pogor}.
For such equations Pogorelov \cite{Pogor} has proved the existence of a
generalized solution in any convex domain on the plane.
Also the Dirichlet problem has been solved and properties of regularity
of
the solution have been investigated.
We cannot directly apply to our case these results because in our case
$g=\sqrt{\phi_x^2 +\phi_y^2}\geq 0$, i.e. it  is positive
but not  strictly
positive and moreover we have a boundary problem which is not of a
Dirichlet type, but rather of the Neumann type (\ref{bcMA}).
We will consider equation (\ref{MoAm}) in another work.

\subsection{Quasi-linear Equation}

Let us now  take a parameterization ("gauge") of the surface
defined by (\ref{1eq}) and (\ref{3eq}) as
\be
K(y_1,y_3)=y_1
\label{K31}
\ee
This parameterization is consistent with the boundary conditions.
One can see that this parameterization is related with condition
(\ref {pro1})
after  a change of
variables $y_1\to y'_1$,
such  that
$
(\partial _1 K)^{-1}\partial / \partial y_1
$ $\to$
$\partial/ \partial y'_1$.

In parameterization (\ref{K31}) we have

\begin{equation}
 \partial _3 f
=-2(2  \partial_{1} g+ g\partial_{1}h);
\label{e1k1}
\end{equation}

\begin{equation}
 \partial _3 g=
2(2  \partial_{1} f-f\partial_{1}h);
\label{e2k1}
\end{equation}

\begin{equation}
 \partial _3 h=
 \partial_{1} (2fg).
\label{e3k1}
\end{equation}
Equation (\ref{e3k1}) can be rewritten in the form
of conservation low   for functions $f$ and $g$

\begin{equation}
 \partial _3 (g^2-f^2)- \partial_{1} (8fg)=0.
\label{cc}
\end{equation}

From this equation follows that there exists a function $\Psi$
such that

\be
fg=\frac{1}{8}\partial _3 \Psi,
\label{psi11}
\ee

\be
g^2-f^2=\partial _1 \Psi.
\label{psi21}
\ee
One can express $g^2$  and $f^2$  in term of $\Psi$ as

\be
g^2=\frac{1}{2}(\partial _1 \Psi +\sqrt{(\partial _1 \Psi )^2
+(1/4\partial _3 \Psi)^2  })
\label{g21}
\ee

\be
f^2=\frac{1}{2}(-\partial _1 \Psi +\sqrt{(\partial _1 \Psi )^2
+(1/4\partial _3 \Psi)^2  })
\label{f21}
\ee

\be
f^2+g^2=\sqrt{(\partial _1 \Psi )^2
+(\frac{1}{4}\partial _3 \Psi)^2 }
\label{g+f1}
\ee

Substituting (\ref{psi11}), (\ref{psi21}) and (\ref{g+f1})
to the equation
\be
\partial_3 (fg)=
2 \partial_1 (f^2-g^2)-
2\partial_1 h (g^2 +f^2)
\ee
which follows from (\ref{e1k1}), (\ref{e1k1}) and (\ref{K31})
one gets the following

{\bf Proposition.}
If $\Psi$ is a solution of

\begin{equation}
\frac{1}{4} \partial ^2 _3 \Psi+4\partial^2 _1
\Psi +(\partial ^2_1\Psi -1)
\cdot
\sqrt{(\frac{1}{4}\partial _3 \Psi )^2
+(\partial_1 \Psi)^2}
=0
\label{epsi}
\end{equation}
then $f$ and $g$
defined by (\ref{f21}) and (\ref{g21}), and
$ h=\frac{1}{4}(g^2-f^2)$
satisfy (\ref{e1k1})-(\ref{e3k1}).

Boundary conditions for equation (\ref{epsi}) are the following
\be
\partial _3\Psi|_{y_3 \to \pm\infty}=\pm 8\zeta;
~~~~
\partial _1\Psi|_{y_3 \to \pm\infty}=a(y_1);
\ee
where $ a(y_1)$ is a solution of the
following equation

\be
ae^{\frac{1}{4}(a-y_1)} + e^{\frac{1}{2}(a-y_1)}
-\zeta ^2=0.
\label{sbc}
\ee

Introducing $\phi(x,y)=\Psi (y_1,2y_3)$ one rewrites
equation (\ref{epsi}) as

\be
\phi_{yy}+ \phi_{xx}
\cdot  (4+\sqrt{\phi _x^2 +\phi _y^2}~)
=
\sqrt{\phi _x^2 +\phi _y^2}.
\label{epsii}
\ee

\section{Conclusion}

In this note we have shown that equations for the domain wall
for $SU(2)$ group can be reduced to the Monge-Amp\`{e}re
equation (\ref{MoAm}) or to equation (\ref{epsii}).
The boundary conditions for these equations
are nonstandard and they require a further investigation.
We also constructed a formal solution for $U(1)$ symmetric domain wall.

\section{Acknowledgments}
I would like to thank
Laboratoire de Physique Theorique et Hautes Energies and especially
L. Baulieu for invitation and kind hospitality
and  acknowledge the support of the CNRS grant.
I am very grateful to J. Sonnenschein for sharing the results
of \cite{KSY} prior to publication and correspondence
and to J. Maldacena for very useful comments.

\bigskip

\smallskip

{\Large {\bf Appendix}}

\bigskip

{\large {\bf A ~~~Algebraic relations}}

\smallskip

We will prove here the algebraic
Kaplunovsky-Sonnenschein-Yankielowicz (KSY) formula  \cite{Scern,KSY}
which is used
in the investigation of the supersymmetric cycles.

We use the following notations:

$A_m$ , $m=1,2,3$ are complex 3-vectors,
$A_m\in C^3$  with components $A_{m\alpha}\in
C$, $\alpha =1,2,3$. The scalar product

\be
(A_m,A_n)=\sum _{\alpha =1}^{3}{\bar A}_{m\alpha}A_{n\alpha},
\label{spod}\ee
here  ${\bar A}_{m\alpha}$ means the complex conjugation.
Note that

\be
(A_n,A_n)\geq 0 , ~~(no~ summation)
\ee

\be
Re(A_m,A_n)=
Re(A_n,A_m),~~~~~~~
Im(A_m,A_n)=-Im(A_n,A_m).
\ee

One has the known formula for the Gram determinant

\be
det|| (A_m,A_n)||= det|| {\bar A}_{m\alpha}||\cdot det|| A_{m\alpha}||,
\label{dgram}
\ee
which follows from the matrix relation

\be
||(A_m,A_n)||=|| \bar{A}_{m\alpha}||\cdot || A^{tr}_{m\alpha}||.
\label{matrel}
\ee

One also has

\be
det||A_{m\alpha}||=\epsilon
_{\alpha\beta\gamma} A_{1\alpha}A_{2\beta}A_{3\gamma}=
 (\bar{A_{3}}, A_{1}\times A_{2}).
 \label{3mat} \ee

Therefore the Gram determinant is
\be
det|| (A_m,A_n)||=| (\bar{A_{1}}, A_{2}\times A_{3})|^2.
\label{gramd}
\ee

One has the following

\bigskip

{\bf Lemma 1.}

$$
det|| (A_m,A_n)||=
det|| Re(A_m,A_n)||-
$$

\be
 (A_1,A_1)Im ^{2}(A_2,A_3)
-(A_2,A_2)Im ^{2}(A_3,A_1)-(A_3,A_3)Im ^{2}(A_1,A_2).
\label{lemma1}
\ee

\bigskip

{\bf Proof.} Let us denote

$$
x_{mn}=Re(A_m,A_n),~~~~~y_{mn}=Im(A_m,A_n).$$
Then we have

$$det||(A_m,A_n)|| =
\sum _{p}(-)^{p}
(x_{1p(1)}+iy_{1p(1)})\cdot
(x_{2p(2)}+iy_{2p(2)})\cdot
(x_{3p(3)}+iy_{3p(3)})=
$$

\be
\sum _{p}(-)^{p}
(x_{1p(1)}x_{2p(2)}x_{3p(3)}-x_{1p(1)}y_{2p(2)}y_{3p(3)}
-x_{2p(2)}y_{1p(1)}y_{3p(3)} -x_{3p(3)}y_{1p(1)}y_{2p(2)})=
\label{l1}
\ee

\be
det||Re(A_m,A_n)||-
\sum _{p}(-)^{p}
(x_{1p(1)}y_{2p(2)}y_{3p(3)}
+x_{2p(2)}y_{1p(1)}y_{3p(3)}
+x_{3p(3)}y_{1p(1)}y_{2p(2)})
\label{l2}
\ee

$$
=det||Re(A_m,A_n)||-
$$

\be
(A_1,A_1)Im ^{2}(A_2,A_3)
-(A_2,A_2)Im ^{2}(A_3,A_1)-(A_3,A_3)Im ^{2}(A_1,A_2).
\label{fl2}
\ee
 In the line (\ref {l1}) we have used that the Gram determinant is real.
To get line (\ref{fl2}) from (\ref{l2}) note that
$y_{11}=y_{22}=y_{33}=0$
and that terms of the form $x_{12}y_{23}y_{32}$ are canceled out due
to the factor $(-)^p$.

\bigskip

{\bf Lemma 2}.

\be
det|| Re(A_m,A_n)||=| (\bar{A}_3,A_1\times A_2)|^2+
|\frac{1}{2}\epsilon_{mnk}A_m Im (A_n,A_k)|^2.
\label{l3}
\ee

\bigskip

{\bf Proof}.
The relation (\ref{l3}) follows from (\ref {lemma1}) due to
(\ref{gramd})   and

$$
|\frac{1}{2}\epsilon _{mnk}A_m Im (A_n,A_k)|^2=
$$

\be
(A_1,A_1)Im ^{2}(A_2,A_3)
+(A_2,A_2)Im ^{2}(A_3,A_1)+
(A_3,A_3)Im ^{2}(A_1,A_2).
\ee

Let us set
\be
 h_{mn}=Re (A_m,A_n)+\delta_{m3}\delta_{n3},~~m,n=1,2,3.
 \label{hmat}
 \ee

\bigskip

{\bf Lemma 3}. One has
\be
det|| h_{mn}||=det|| Re(A_m,A_n)|| +|A_1\times A_2|^2-Im^2(A_1,A_2).
\label{lemma3}
\ee
\bigskip

{\bf Proof.}
The matrix $||h_{mn}||$ has the form:
\begin{equation}
||h_{mn}||=\left(
\begin{array}{ccc}
  (A_1,A_1)
   &Re (A_1,A_2) & Re (A_1,A_3) \\
 Re (A_2,A_1)&(A_2,A_2)&Re (A_2,A_3) \\
 Re (A_3,A_1)&Re (A_3,A_2)&(A_3,A_3)+1
 \end{array}
\right)
\end{equation}
The determinant of $||h_{mn}||$ can be represented as

$$
det \left(
\begin{array}{ccc}
 (A_1,A_1)&Re (A_1,A_2) & Re (A_1,A_3) \\
 Re (A_2,A_1)& (A_2,A_2)&Re (A_2,A_3) \\
 Re (A_3,A_1)& Re (A_3,A_2) & (A_3,A_3)
\end{array}
\right)
+det \left(
\begin{array}{ccc}
 (A_1,A_1)&Re (A_1,A_2) & 0 \\
 Re (A_2,A_1)& (A_2,A_2)&0 \\
 0 & 0 & 1
\end{array}
\right)=
$$

\be
=det||Re(A_m,A_n)||+(A_1,A_1)(A_2,A_2)-Re^2(A_1,A_2).
\label{sumdet}
\ee
Using the identities
\be
|A_1\times A_2|^2=
(A_1\times A_2,A_1\times A_2) =
(A_1,A_1)(A_2,A_2)-|(A_1,A_2)|^2,
\label{sqv}
\ee

\be
|(A_1, A_2)|^2 =Re^2 (A_1,A_2)+ Im^2 (A_1,A_2)
\label{sqv'}
\ee
we get the proof of Lemma 3.

One has the following basic algebraic formula \cite{Scern,KSY}

\bigskip

{\bf Proposition} (V. Kaplunovsky, J. Sonnenschein and  S. Yankielowicz).
One has the following representation
for the determinant of the matrix $||h_{mn}||$:
\be
det|| h_{mn}||=Im^2[(A_1,A_2)- (\bar{A_{3}}, A_{1}\times A_{2}) ]
+Re^2(\bar{A}_{3}, A_{1}\times A_{2}) +
\label{format}
\ee
$$
+|A_1\times A_2
+i\frac{1}{2}\epsilon_{mnk}{\bar A}_m Im (A_n,A_k)|^2.
$$
\bigskip
{\bf Proof} of  the KSY formula follows from Lemmas 1,2 and 3.

Let us set

\be
\Phi =Im [(A_1,A_2)- (\bar{A_{3}}, A_{1}\times A_{2})]
 \label{phi}
\ee

\bigskip

{\bf Corollary } (V.Kaplunovsky, J.Sonnenschein and S.Yankielowicz).
The
relation \be det|| h_{mn}||=\Phi^2 \label{theo21} \ee is equivalent to \be
A_1\times A_2
+i\frac{1}{2}\epsilon_{mnk}{\bar A}_m Im (A_n,A_k)=0
\label{theo22}
\ee
{\bf Proof.} From the KSY formula (\ref{format}) it follows
that to prove the
Corollary one has to show that the relation

\be
A_1\times A_2
+i\bar{A}_1 Im (A_2,A_3)
+i\bar{A}_2 Im (A_3,A_1)
+i\bar{A}_3 Im (A_1,A_2)=0
\label{dr}
\ee
implies the equality

\be
Re(\bar{A}_3,A_1\times A_2)=0.
\label{Req}
\ee
Let us take  the scalar product of (\ref{dr}) with $\bar {A}_3$.  Then
we get

\be
(\bar {A}_3, A_1\times A_2)
+i(\bar{A}_3 ,\bar{A}_1 )Im (A_2,A_3)
+i(\bar{A}_3 ,\bar{A}_2 )Im (A_3,A_1)
+i(\bar{A}_3 ,\bar{A}_3 )Im (A_1,A_2)=0,
\label{sdr}
\ee
or

\be
(\bar {A}_3, A_1\times A_2)
+i(A_1 ,A_3 )Im (A_2,A_3)
+i(A_2 ,A_3 )Im (A_3,A_1)
+i(A_3 ,A_3 )Im (A_1,A_2)=0.
\label{csdr}
\ee
Now  let us take the real part of (\ref{csdr})

\be
Re(\bar {A}_3, A_1\times A_2)=
 Im(A_1 ,A_3 )Im (A_2,A_3)
+Im(A_2 ,A_3 )Im (A_3,A_1).
\label{icsdr}
\ee
The right hand side  of (\ref{icsdr}) vanishes since
$Im(A_2 ,A_3 )=-Im(A_3 ,A_2 )$.
The Corollary is proved.

Let us  notice that there is also an equivalent  formulation of
the  KSY proposition and Corollary, that can be formulated as

\bigskip
{\bf Theorem}. One has the following relation

\be
det ||h_{mn}||=\Phi^2 +|R|^2 +Re^2({\bar A}_3,R),
\label{Theorem}
\ee
where

\be
R=A_1\times A_2 +\frac{i}{2}\epsilon_{mnk}{\bar A}_m Im(A_n,A_k).
\label{R=A}
\ee

{\bf Proof.} The theorem follows from Lemmas 1,2,3 and the relation

\be
Re({\bar A}_3,A_1\times A_2)=
Re({\bar A}_3,R).
\label{proof}
\ee

This formulation is convenient because
the corollary now is more clear since

\be
|R|^2 +Re^2({\bar A}_3,R)=0
\ee
is obviously equivalent to  $R=0$.


\begin{thebibliography}{99}


\bibitem{HW} A. Hanany and E. Witten. ``Type IIB Superstrings, BPS
Monopoles,
and Three-Dimensional Gauge Dynamics, Nucl. Phys.{\bf B492}(1997)152.

\bibitem{Witten1} E. Witten, ``Solutions Of Four-Dimensional Field
Theories From M-Theory'', Nucl. Phys.{\bf B500}(1997)3.
hep-th/9703166.

\bibitem{EGK} S. Elitzur, A. Giveon, D. Kutasov. ``Branes and $N=1$
Duality in String Theory,'' Phys. Lett.{\bf B400}(1997)269.

\bibitem{BersVafa} M. Bershadsky, A. Johansen, T. Pantev, V. Sadov
and  C. Vafa, "F-theory, Geometric Engineering and N=1 Dualities"
Nucl.Phys. {\bf B505} (1997) 153-164,

\bibitem{OV} H. Ooguri and C. Vafa.''Geometry of $N=1$
Dualities in Four Dimensions'', Nucl. Phys. {\bf B500}(1997)62,
hep-th/9702180.

\bibitem{EJS} N. Evans. C.V. Johnson, A. D Shapere. ``Orientifolds,
Branes And Duality of 4D Gauge Theories'', Nucl. Phys. {\bf 505}
(1997)251, hep-th/9703210.

\bibitem{BH} J. Brodie and A. Hanany, ``Type IIA Superstrings, Chiral
Symmetry, and
$N=1$ 4D Gauge Theory Dualities'', Nucl.Phys. {\bf B506}(1997)157.
hep-th/9704043.

\bibitem{BSTY} A. Brandhuber, J. Sonnenschein, S. Theisen, and S.
Yankielowicz,
``Brane Configurations And 4D Field Theory Dualities'',
Nucl.Phys. {\bf B502}(1997)125.
hep-th/9704044.

\bibitem{T} R. Tatar, ``Dualities In 4D theories With Product
Gauge Groups From Brane Configurations'', hep-th/9704198.

\bibitem{Witten} E. Witten, ''Branes And The Dynamics Of QCD'',
Nucl.Phys. {\bf B507}(1997)658.
hep-th/9706109.

\bibitem{HOO} K. Hori, H. Ooguri, and Z. Oz, ``Strong Coupling Dynamics of
Four-Dimensional $N=1$ Gauge Theory from M Theory Five brane'',
hep-th/9706082.

\bibitem{HSZ} A.Hanany, M.Strassler and A.Zaffaroni, "Confinement
and Strings in MQCD", hep-th/9707244.

\bibitem{HOO2} Jan de Boer, Kentaro Hori, Hirosi Ooguri and  Yaron Oz
 "Kahler Potential and Higher Derivative Terms from M
Theory Fivebrane", hep-th/9711143.

\bibitem{BIKSY} A. Brandhuber, N. Itzaki. V.. Kaplunovsky, J. Sonnenschein,
and
S. Yankielowicz, ``Comments On The M Theory Approach to $N=1$ SQCD And
Brane
Dynamics,''
Phys.Lett. {\bf 410}(1997)27.
hep-th/9706127.

\bibitem{Sin} S.Nam, K. Oh and S.-J.Sin, "Super-potentials
of $N=1$ Supersymmetric Gauge Theories from M-theory",
hep-th/9707247.

\bibitem{AOT} C.Ahn, K.Oh and R.Tatar, "M Theory Fivebrane
Interpretation for
Strong Coupling Dynamics of $SO(N_c)$ Gauge Theories",
Phys.Lett. {\bf B416}(1998)75.
hep-th/9709096.

\bibitem{Boer} J.de Boer, K.Hori,
H.Ooguri and Y.Oz, ``Branes and Dynamical Supersymmetry Breaking,''
hep-th/9801060.

\bibitem{itep} A. Marshakov, M. Marellini, and A. Morozov, ``Insights
and Puzzles from Branes: 4d SUSY Yang-Mills From 6d Models'',
hep-th/9706050.

\bibitem{Becker1} K. Becker, M. Becker and A. Strominger.
``Fivebranes, membranes and nonperturpative string theory'',
Nucl. Phys.{\bf B456}(1995) 130.

\bibitem{Becker} K. Becker, M. Becker, D. R. Morrison, H. Ooguri,
Y. Oz, Z. Yin.
``Supersymmetric Cycles in Exceptional Holonomy Manifolds and
Calabi-Yau 4-folds,'' Nucl. Phys.{\bf B480}(1996) 225.

\bibitem{Shifman} G. Dvali and M. Shifman. ``Domain Walls In Strongly
Coupled Theories,''
Phys.Lett. {\bf 396}(1997)64.
hep-th/9612128.

\bibitem{KSS} A. Kovner, M. Shifman, and A. Smilga,
``Domain Walls in Supersymmetric Yang-Mills Theories,''
Phys.Rev. {\bf D56}(1997)7978.
hep-th/9706089.

\bibitem{AV} A. Volovich. ``Domain Wall in MQCD and
Supersymmetric Cycles in Exceptional Holonomy Manifolds,''
hep-th/9710120.

\bibitem{Joyce} D. Joyce.  ``Compact Riemannian 7-Manifolds
with Holonomy $G_{2}$,'' I;II, J. Diff. Geom.(1996){\bf 43} p. 291; 329.

\bibitem{SandC} F. R. Harvey, ``Spinors and Calibrations,'' Academic Press
(1990).

\bibitem{Vafa} S. L. Shatashvili and C. Vafa. ``Superstrings
and Manifolds of Exceptional Holonomy,'' hep-th/9407025.

\bibitem{Fayya} A. Fayyazuddin and M. Spalinski. ``Extended Objects in
MQCD,''
hep-th/9711083.
\bibitem{Scern} J. Sonnenschein, Talk on the 3rd Workshop on recent
developments in Theoretical physics,
CERN, December, 1997.

\bibitem{KSY} V. Kaplunovsky, J. Sonnenschein and  S. Yankielowicz,
in preparation.

\bibitem{BKS1} L. Baulieu, H. Kanno and  I. M. Singer,
Cohomological Yang-Mills Theory in Eight Dimensions,
hep-th/9705127.

\bibitem{BKS2} L. Baulieu,  H. Kanno and I. M. Singer,
Special Quantum Field Theories In Eight And Other Dimensions,
hep-th/9704167.

\bibitem{Pogor} A.V.Pogorelov, External Geometry of Convex
Surfaces.Translations of Mathematical Monographs, v.35. AMS, 1973.
Monge-Amp\`{e}re Equations of Elliptic Type, Groninger,  Netherlands,
1964.

\bibitem{TA} Thierry Aubin,  Nonlinear Analysis on Manifolds.
Monge-Amp\`{e}re Equations, Springer-Verlag,  New York, Heidelberg,
Berlin, 1982.

\bibitem{Yau} S-T. Yau, On the Ricci curvature of a compact K\"{a}hler
manifold and the complex
Monge-Amp\`{e}re equation I. Comm. Pure Appl. Math. 31 (1978) 339-411.

\end{thebibliography}
\end{document}